\documentclass[12pt]{article}
\pdfoutput=1
\usepackage{amsmath}
\usepackage[hmargin=1in,vmargin=1in]{geometry}
  \usepackage{paralist}
  \usepackage{graphics,ulem} 
  \usepackage{epsfig} 
\usepackage{graphicx}
\usepackage{epstopdf} 
 \usepackage[colorlinks=true]{hyperref}
\hypersetup{urlcolor=blue, citecolor=red}
\usepackage{array}

\begin{document}

\title{Limiting the Development of Anti-Cancer Drug Resistance in a Spatial Model of Micrometastases}

\maketitle


\centerline{\scshape Ami B. Shah }

{\footnotesize
\centerline{Department of Biology}
\centerline{The College of New Jersey}
\centerline{Ewing, NJ, USA}
} 

\medskip

\centerline{\scshape Katarzyna A. Rejniak }
{\footnotesize
\centerline{Integrated Mathematical Oncology Department}
\centerline{and Center of Excellence in Cancer Imaging and Technology}
\centerline{H. Lee Moffitt Cancer Center and Research Institute,}
\centerline{Department of Oncologic Sciences}
\centerline{University of South Florida}
\centerline{Tampa, FL, USA}
} 

\medskip

\centerline{\scshape Jana L. Gevertz}
{\footnotesize
\centerline{Department of Mathematics and Statistics}
\centerline{The College of New Jersey}
\centerline{Ewing, NJ, USA}
} 

\bigskip

\begin{abstract}
While chemoresistance in primary tumors is well-studied, much less is known about the influence of systemic chemotherapy on the development of drug resistance at metastatic sites. In this work, we use a hybrid spatial model of tumor response to a DNA damaging drug to study how the development of chemoresistance in micrometastases depends on the drug dosing schedule. We separately consider cell populations that harbor pre-existing resistance to the drug, and those that acquire resistance during the course of treatment. For each of these independent scenarios, we consider one hypothetical cell line that is responsive to metronomic chemotherapy, and another that with high probability cannot be eradicated by a metronomic protocol. Motivated by experimental work on ovarian cancer xenografts, we consider all possible combinations of a one week treatment protocol, repeated for three weeks, and constrained by the total weekly drug dose. Simulations reveal a small number of fractionated-dose protocols that are at least as effective as metronomic therapy in eradicating micrometastases with acquired resistance (weak or strong), while also being at least as effective on those that harbor weakly pre-existing resistant cells. Given the responsiveness of very different theoretical cell lines to these few fractionated-dose protocols, these may represent more effective ways to schedule chemotherapy with the goal of limiting metastatic tumor progression. 
\end{abstract}

\newpage
\section{Introduction}

\subsection{Commonly Used Chemotherapeutic Schedules}
The standard cancer chemotherapy treatment protocol is to give a drug at the maximum tolerated dose (MTD), with rest periods in between \cite{Ledzewicz13}. In theory, treating with the MTD is meant to elicit the strongest tumor killing possible without inducing intolerable levels of toxicity. MTD chemotherapy has been successful in the treatment of many hematological cancers, but this approach has failed to demonstrate sustained responses in the majority of solid tumors \cite{Malik14}. 

As an example in which MTD is routinely used, the treatment of ovarian cancer typically involves surgical debulking followed by cycles of intermittent chemotherapy: systemic exposure to high doses of chemotherapeutics is followed by 3-4 week treatment-free intervals to allow for recovery of healthy tissue \cite{DeSouza11}. Although this is the standard treatment protocol, clinical trials involving ovarian cancer patients have demonstrated that there is a survival benefit for decreasing the length of the intervals between chemotherapy treatments from 3 weeks to 1 week \cite{DeSouza11}. Consistent with this finding, preclinical trials have demonstrated that continuous drug exposure results in greater antitumor efficacy than an intermittent schedule in ovarian cancer xenografts due to greater tumor cell kill, reduced proliferation, and reduced angiogenesis \cite{DeSouza11}. Whether considering the case of continuous infusion, or considering decreasing the time off between chemotherapy doses, drug needs to be administered at a lower dose than the MTD to avoid the accumulation of toxicity. 

The administration of a chemotherapeutic agent at a relatively low, minimally toxic dose, for extended periods of time with no prolonged drug-free break is referred to as metronomic chemotherapy (MC) \cite{HanahanEtAl:2000}. MC has proven especially effective as an anti-angiogenic treatment, in which the chemotherapeutic agents are directed towards the killing of the endothelial cells, inhibiting their migration, or towards decreasing activity of endothelial progenitor cells in the bone marrow \cite{HanahanEtAl:2000,MrossSteinbild:2012,ScharovskyEtAl:2009}. Several clinical trials using this approach are currently active \cite{MrossSteinbild:2012}. In addition, MC is used as a maintenance treatment to prevent tumor progression after initial application of MTD therapy halts tumor growth \cite{EmmeneggerEtAl:2011,VivesEtAl:2013}. 

It is also hypothesized that one of the advantages MC has over the MTD approach in solid tumors is that it limits the emergence of drug resistant cancer cells. Drug resistance clinically manifests when a patient initially responds well to chemotherapy, but becomes unresponsive to the protocol after multiple cycles \cite{Baguley:2010,DeSouza11,KimTannock:2005,ZahreddineBorden:2013}. Unfortunately, such drug resistance is quite common in solid tumors, and drug resistance is the leading cause of chemotherapy failure in the treatment of cancer \cite{DeSouza11}. Both extrinsic (environmental) and intrinsic (molecular) factors can induce drug resistance, and drug resistance can be a pre-existing phenomenon, acquired phenomenon, or both. Pre-existing (primary) drug resistance occurs when the tumor contains a subpopulation of drug resistant cells at the initiation of treatment, and these cells are selected for during the course of therapy. Acquired (emerging) resistance involves the adaptation of a tumor cell subpopulation so that the cells gradually develop drug resistance \cite{Tredan07, Wu13}.

\subsection{Related Mathematical Models}
Mathematical modeling is now widely used to predict anti-cancer drug efficacy, to better understand mechanisms of drug resistance, and to identify novel treatment protocols that limit the emergence of resistance (see \cite{Brocato14, Foo14, Lavi12} for excellent reviews of modeling work done in these areas). Pioneering work of Coldman and Goldie (e.g., see \cite{Coldman86}) studied the stochastic switching of cells between sensitive and resistant compartments as a result of point mutations, and uses mathematical predictions to guide treatment schedules. More advanced stochastic models have been developed \cite{Bozic13,Foo13,Foo09, Foo10, Komarova05, Komarova07, Mumenthaler11}, and these have been used to explore a range of questions. For example, Komarova and Wodarz developed a model for multi-drug resistance using a multi-type birth-death process. In these models, resistance to each targeted small-molecule drug was conferred by a single random genetic mutation, and they explore how the number of drugs needed to prevent treatment failure depends on the initial size of the tumor at the start of treatment, as well as the role of cell quiescence \cite{Komarova05, Komarova07}. As another example, Foo and Michor used a nonhomogeneous multi-type birth-death process model to investigate optimal dosing of a targeted anti-cancer drug to minimize the risk of resistance (assuming resistance is due to a single (epi)genetic alteration) constrained by drug toxicity \cite{Foo09, Foo10}. This work was later extended to consider combination therapy with a small molecule inhibitor and a chemotherapeutic agent \cite{Mumenthaler11}. 

Non-spatial continuum-based approaches have also been widely used to model drug resistance and to try and control its emergence. Optimal control problems have been formulated to analyze nonlinear ordinary differential equation models involving resistant and sensitive cancer cells (see, e.g., \cite{Hadjiandreou13, LedzewiczEtAl:2015,Ledzewicz06}). As an example, Ledzewicz and Sch\"{a}ttler studied how bang-bang controls (full doses with rest periods in between) and singular controls (time-varying partial doses) could be used to find chemotherapy schedules that control and prolong the onset of drug resistance \cite{Ledzewicz06}. This work was recently extended to include metronomic therapy \cite{LedzewiczEtAl:2015}. Also using a non-spatial continuum approach, Gatenby and colleagues explore tumor dynamics and control in response to an evolutionary ``double-bind" therapy in which cellular adaptations to one treatment renders the cancer increasingly vulnerable to a second therapeutic attack \cite{Cunningham11, Orlando12}. Unlike in \cite{Ledzewicz06} and \cite{Hadjiandreou13}, a continuous spectrum of phenotypic resistance is considered in these models. A continuous spectrum of phenotypic resistance has also been considered by Lorz et al. using an integro-differential equation framework to describe mutations between sensitive (but more fit) and resistant (less fit) cancer cells \cite{Lorz13}. Levy and colleagues built upon this work to include the effects of density dependence on cancer cell division and death rates, as well as to account for both genetic and epigenetic changes that lead to resistance \cite{Greene14,Lavi13}. 

While interesting insights have been drawn from stochastic branching models and deterministic continuum models, both approaches neglect to consider the important role played by spatial (microenvironmental) heterogeneity in the formation of drug resistance. Spatial impacts on the dynamics of pre-existing drug resistance were first considered by Jackson and Byrne using partial differential equations to study spherically symmetric growth and response to treatment (under bolus versus continuous injection) of a heterogeneous tumor composed of two cell types: low proliferation but high drug resistance or high proliferation but more drug sensitive \cite{Jackson00}. Nowak and colleagues considered treatment with a targeted therapy in a non-symmetric three-dimensional spatial model of heterogeneous tumor progression at metastatic sites. Assuming drug is uniformly distributed in the tumor, they quantified how the temporal probability of regrowth varies depending on the net growth rate of the tumor and the dispersal probabilities \cite{Waclaw15}. Finally, the work of Lorz et al. \cite{Lorz13} was also extended to include diffusion of drug and nutrient in a spherically symmetric environment. By including a spatial component, they could study how tumor cell heterogeneity and resistance adapt to the surrounding environment \cite{Lorz15}. In each of these spatial works, a uniform or spherically symmetric microenvironment is assumed, which over-simplifies much of the microenvironmental heterogeneity that influences the emergence of drug resistance.

There have only been a handful of attempts to consider the emergence of drug resistance in heterogeneous microenvironments. Silva and Gatenby \cite{Silva10} introduced a two-dimensional cellular automaton model composed of chemosensitive cells that are rapidly dividing and chemoresistant cells that are slowly dividing. Accounting for spatial distributions of oxygen, glucose and pH buffers, they found that optimal tumor control in the case of pre-existing resistance is achieved when a glucose competitor drug was followed by cytotoxic chemotherapy in two separate doses \cite{Silva10}. Menchon developed a two-population (resistant and sensitive), two-dimensional cellular automaton model and used it to study how pre-existing resistance and acquired resistance that is not drug-induced influences treatment with an otherwise successful chemotherapeutic drug \cite{Menchon15}. Powathil et al. developed a two-dimensional hybrid multiscale cellular automaton model to study the effects of cell-cycle based chemotherapeutic drugs on cancer cell populations with drug resistance in heterogeneous microenvironments \cite{Powathil}.

In contrast to the previous mathematical models that consider anti-mitotic drugs and assume that chemoresistant cells are proliferating less frequently, we developed a hybrid spatial off-lattice model to explore the resistance dynamics induced by exposure to a DNA damaging drug \cite{Gevertz15}. Unlike previous models, in this work the impact of the heterogeneous microenvironment in the case of both pre-existing and acquired resistance is quantified. We found that microenvironmental pressures modeled through a heterogeneous spatial configuration of blood vessels does not significantly impact transient and long-term tumor behavior when resistance is pre-existing, but does have a significant impact when resistance is acquired. In fact, in the case of acquired resistance, we showed that microenvironmental niches of low drug/sufficient oxygen and low drug/low oxygen play an important role in the development of resistance \cite{Gevertz15}. The conclusion that acquired resistance is likely to arise first in the ``poor drug penetration sanctuaries" was recently independently established in \cite{Fu15} using a multi-compartment model of an anti-mitotic drug in which chemoresistant cells have a growth disadvantage. Unlike in our study \cite{Gevertz15}, the work in \cite{Fu15} assumes spatial heterogeneity between different metastatic sites, but not within a metastatic site. That said, it is the fact that both models account for spatial heterogeneity (explicitly or implicitly) that allow for such a prediction to be made.

\subsection{Summary of Current Work}
Focusing on the influence of systemic che\-mo\-therapy at metastatic sites, we point out that relatively less attention has been spent investigating this question as compared to studying drug impact at the site of the primary tumor. Experimentally, it is difficult to recreate the whole metastatic cascade starting with cancer cell invasion from the primary tumor, to vascular transport, to colonization of distant sites \cite{SaxenaChristofori:2013}. The early stages of metastatic development (micrometastases) are also difficult to monitor since small masses of tumor cells are clinically undetectable. Moreover, the use of mouse models is limited in time as the animals must be euthanized once their primary tumors reach a certain size. 

There is also limited mathematical literature devoted to metastatic response to systemic chemotherapy \cite{BenzekryHahnfeldt:2013,CunninghamEtAl:2015,Fu15}. In the present work, we build off our previously developed self-calibrated spatial hybrid model  \cite{Gevertz15} to explore the effects of DNA damaging drug dosing schedule (MTD, metronomic, fractionated) on the development of chemoresistance in simulated tumor micrometastases. We show that in the case of strongly acquired resistance, micrometastases which, with high probability, become resistant to a metronomic chemotherapy can actually be eradicated with high probability by a small number of fractionated-dose (FD) protocols \cite{Buffoni06}, most of which include off-days in the later half of the week. A subset of these FD protocols are also at least as effective as MC at eradicating weakly resistant (whether acquired or pre-existing) micrometastases, which are defined as those that can be eradicated by metronomic therapy. On the other hand, if the pre-existing resistance to the drug is so strong that tumor does not respond to metronomic therapy, it is more challenging to design a treatment protocol to overcome this resistance.

\section{Mathematical Model}

\begin{figure}[ht!]
 \begin{center}
  \includegraphics[width=0.95\textwidth]{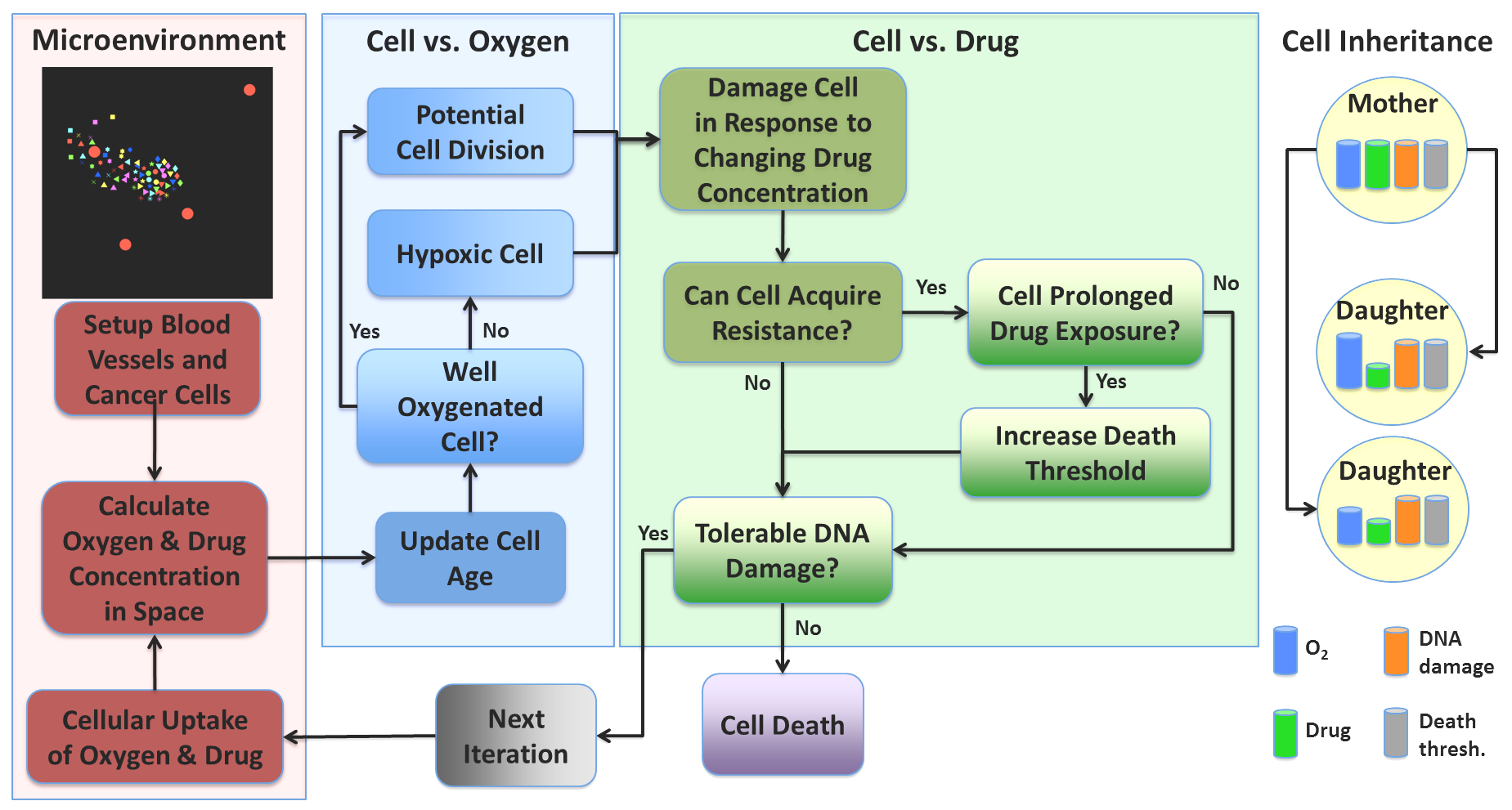}\\
  \caption{\footnotesize{Flowchart of cell behavior in response to microenvironmental factors of oxygen and drug. All simulations start with the same initial configuration of blood vessels and cancer cells (top of first column, vessels represented by red circles). Based on signals sensed from the microenvironment (first column), each cell can respond to oxygen levels (second column) by potentially proliferating or becoming hypoxic. Cells also respond to drug levels (third column) by either surviving, adapting or dying. Upon cell division (rightmost column), some cell properties are inherited by daughter cells (DNA damage and death threshold), whereas others are determined by the microenvironment (oxygen and drug uptake). Adapted from \cite{Gevertz15}}.}
  \label{figure:flowchart}
  \end{center}
\end{figure}

Previously, we developed a hybrid off-lattice discrete-continuous model of a two-dimensional tissue slice in which a tumor grows, interacts with its microenvironment, and is treated with a DNA damaging chemotherapeutic drug \cite{Gevertz15}. Within the tissue slice, the position of four non-evolving blood vessels is imposed, and these act as the source of both oxygen  and drug. Gradients of oxygen and drug in tissue space, determined by numerically solving the associated partial differential equations (PDEs), impact the behavior of the discrete cancer cells in the model. Cells will either remain viable and proliferate or become hypoxic and quiescent based on both oxygen levels (cell-microenvironment interactions) and the crowdedness of the local microenvironment (cell-cell interactions via a particle-spring framework). Additionally, cells are damaged based on how much of the DNA damaging drug they uptake at each discrete time point, and cancer cells can die if their damage level exceeds the maximum amount of damage tolerable by that cell. The clonal predecessor (one of the initial 65 tumor cells) of each surviving cell is also tracked in order to determine which clones of origin, if any, survive treatment with the DNA damaging drug \cite{Gevertz15}. The entire algorithm indicating cellular response to microenvironmental factors in the model is illustrated in Figure \ref{figure:flowchart}. The details of the model are presented in \cite{Gevertz15}, and here we only summarize the discrete and continuous components of the model, as well as the impact of drug action. The small changes between the original model and the model considered herein will be highlighted.


\subsection{Discrete Model Components}

The discrete portion of the model describes cells using an off-lattice, agent-based approach. Each cell is defined by the following: (1) Cell position (determined by cell nucleus, all cells have a fixed radius); (2) Cell age; (3) Cell maturation age (age at which cell is ready to divide); (4) Level of oxygen sensed from local microenvironment, as defined in the PDE for oxygen (determines if cell has become hypoxic); (5) Intracellular drug level, as defined in the PDE for drug  (determined by uptake from local microenvironment); (6) Level of accumulated DNA damage; (7) Level of DNA damage the cell can tolerate (death threshold); (8) Clonal predecessors (cells arising from each of the 65 initial cells). One additional variable has been introduced in this work: (9) Time cell has remained in a continuous hypoxic state. Cells that remain in a sustained state of hypoxia for approximately one day are assumed to die and are removed from tissue space. This modification to the model greatly facilitates comparing the efficacy of multiple treatment protocols, as we do not have to be concerned with the phenomenon seen in \cite{Gevertz15} that some treatments leave behind a handful of completely inert hypoxic cells in low-drug, low-oxygen niches.

The equations of cell mechanics are based on the previously published model by Meineke et al. \cite{Meineke01}, in which cells of a fixed radius (defined by the location of their nucleus) exert Hookean repulsive forces on any cells they come into contact with. A cell responds to the repulsive forces acting upon it by assuming that the connecting springs are overdamped. This causes the system to return to equilibrium without oscillations, and allows us to describe cell velocity as proportional to this repulsive force. More details can be found in \cite{Gevertz15}.



\subsection{Continuous Model Components}

The change in both oxygen $\xi$ and drug $\gamma$ at location $\mathbf{x}=(x,y)$ in the tissue space depends on supply from the vasculature nearby $V_j$ (where $\chi$ is a characteristic function defining the vessel neighborhood), diffusion, and cellular uptake by nearby tumor cells $C_k$. Drug, but not oxygen, also decays in the tissue space. The uncoupled PDEs describing these spatial-temporal dynamics are given by: 
\begin{small}
\begin{equation}
\displaystyle\frac{\partial \xi(\mathbf{x},t)}{\partial t}=\underbrace{\mathcal{D_{\xi}}\Delta \xi(\mathbf{x},t)}_{diffusion} - \underbrace{\min\left(\xi(\mathbf{x},t), \rho_{\xi}\displaystyle\sum_k \chi_{C_k}(\mathbf{x},t)\right)}_{uptake \;\; by \;\; cells} + \underbrace{S_{\xi} \displaystyle\sum_j\chi_{V_j}(\mathbf{x},t)}_{supply},
\end{equation}
\end{small}
\begin{small}
\begin{equation}
\displaystyle\frac{\partial \gamma(\mathbf{x},t)}{\partial t}=\underbrace{\mathcal{D_{\gamma}}\Delta \gamma(\mathbf{x},t)}_{diffusion} -\underbrace {d_{\gamma} \gamma(\mathbf{x},t)}_{decay}-\underbrace{\rho_{\gamma}\displaystyle\sum_k \gamma(\mathbf{x},t)\chi_{C_k}(\mathbf{x},t)}_{uptake \;\; by \;\; cells}+\underbrace{S_{\gamma}(t)\displaystyle\sum_j \chi_{V_j}(\mathbf{x},t)}_{supply}.
\end{equation}
\end{small}

Based on experimental work of cells grown in high and low concentrations of oxygen and glucose \cite{Casciari91,Freyer85}, the uptake term for oxygen is described using zeroth-order kinetics. In particular, the work in \cite{Casciari91,Freyer85} demonstrated that cells grown at high concentrations consume constant amounts of oxygen and glucose that allow them to function normally, while cells grown at low concentrations consume sub-optimal amounts that allow them to survive. Since in our model cells are subjected to gradients of oxygen with very high concentrations near the vasculature, we model oxygen consumption as constant (zeroth-order kinetics with a non-dimensionalized uptake parameter of $\rho_{\xi}=5\times 10^{-5}$).

On the other hand, there are several possible mechanisms of drug uptake. Small molecules can be transported through the cell membrane via pumps -- this will depend on the concentration differences between inner and outer drug concentrations. Targeted drugs are taken up via cell receptor binding -- this will depend on the number of receptors available and whether or not they can be recycled back to the membrane. To simplify these processes we have chosen to model drug uptake using the standard assumption of first-order kinetics (with a non-dimensionalized uptake parameter of $\rho_{\gamma} = 5\times10^{-3}$), contrary to our previous work in \cite{Gevertz15} which used zeroth-order kinetics. 

Initially, a steady-state distribution of oxygen is determined in space in the absence of any cancer cells, and there is only drug at the blood vessel sites. Sink-like boundary conditions are imposed along all domain boundaries, in which the change in concentration with respect to the outward facing normal is proportional to the concentration along the boundary. The PDEs are numerically solved using a forward-difference approximation in time on a square grid (centered difference in space). The PDE parameters, together with a self-calibration process, can be found in \cite{Gevertz15}.

\subsection{Modeling Cell Response to DNA Damaging Drug} 

Each cell has a unique, molecularly-wired sensitivity to drug-induced DNA damage. The cell-level variable that controls this sensitivity in the model is $C_k^{death}$, the maximum amount of drug-induced damage the $k^{th}$ cell can tolerate; we will refer to this as the death threshold of the cell. The duration and level of drug exposure within a cell will determine the increase of DNA damage within the cell ($C_k^{dam}$), while the rate of DNA damage repair ($p$) will regulate the natural decrease of DNA damage. A cell with absolutely no tolerance to the drug ($C_k^{death} = 0$) would die as soon as the slightest bit of drug-induced DNA damage occurs, which is not biologically realistic. Therefore, we assume that in the baseline case of a homogeneous population of chemosensitive cells (i.e., no pre-existing resistance), each cell has a fixed death threshold of $C_k^{death} = Thr_{death} = 0.5$, where the value of $Thr_{death}$ has previously been determined through the parameter self-calibration process \cite{Gevertz15}. 

In the absence of any drug resistance in an initially homogeneous population of cells, the only variation between cells is in how much damage they have accumulated, and this is a function of both space and time. The drug-induced DNA damage is assumed to depend on the current change in intracellular drug concentration ($\Delta \gamma =$ drug uptake minus decay) and on the rate of DNA repair. As in \cite{Gevertz15}, we assume that DNA repair is proportional to the current amount of DNA damage. However, unlike in \cite{Gevertz15}, we do not assume that damage increases at a rate proportional to $\Delta \gamma$, as frontloading the tumor with very high doses of drug would cause such high damage levels that the tumor can be killed with one very large chemotherapy dose. Although this could be true clinically such a dose would not be attainable due to toxicity, and this feature of the model would interfere with finding optimal treatment protocols. Therefore in this model we define
\begin{equation}
 C_k^{dam}(t+\Delta t) = C_k^{dam}(t) + \frac{K_{max}\Delta \gamma}{k_n+\Delta \gamma}\Delta t -  pC_k^{dam}(t)\Delta t.
\label{eqn:damage}
\end{equation}

\noindent In order to ensure consistency with the predictions made at low drug levels (as analyzed in \cite{Gevertz15}), we calibrated model parameters and found that the best fit was achieved for $K_{max} = 8\times10^{-4}$ and $k_n \approx 2.195\times 10^{-4}$. Just as in \cite{Gevertz15}, when the intra\-cellular damage level $C_k^{dam}$ exceeds the death threshold $C_k^{death}$, the cell dies.

In the case of pre-existing drug resistance, a fixed number of cells will be less sensitive to the DNA damaging drug, meaning they can tolerate more damage than the baseline case. These resistant cells have a death threshold of $C_k^{death} = Thr_{multi}\times Thr_{death}$, where $Thr_{multi}>1$ is the pre-existing resistance parameter in the model. The larger $Thr_{multi}$ is, the more resistant this subpopulation of cells is compared to the chemosensitive subpopulation. 

The modeled mechanism of acquired resistance is based on the relatively novel discovery that the stress imposed by cancer-targeting drugs on tumor cells results in (inheritable) phenotypic plasticity -- changes in cell phenotype in the absence of mutations \cite{Pisco15}. In particular, under stress cancer cells have been observed to switch to a stem-like phenotype, making them more resilient in the face of DNA damage \cite{Pisco15}. In our model, it is the sustained exposure to stress from the DNA damaging drug that induces a change in phenotype (further resistance to the drug). To detail, in the case of acquired resistance, the death threshold of each cell $C_k^{death}$ increases independently at the rate $\Delta_{death}$ if the prolonged drug exposure criterion is met:
\begin{equation}
C_k^{death}(t+\Delta t) = 
\left\{ \begin{array}{ll} 
C_k^{death}(t) + \Delta_{death}\Delta t & \mbox{ if } C_k^{exp}(t) > t_{exp} \\
C_k^{death}(t) & \mbox{ otherwise.}\\
\end{array}\right.
\end{equation}

\noindent In this equation, $C_k^{exp}(t)$ keeps track of how long the $k^{th}$ cell has been exposed to significantly high drug concentrations (above $\gamma_{exp}$). Once that exposure time is greater than the threshold time of $t_{exp}$, the cell increases the amount of damage it can tolerate (death threshold) at the rate $\Delta_{death}$. 

\begin{figure}[h!]
 \begin{center}
  \includegraphics[width=0.7\textwidth]{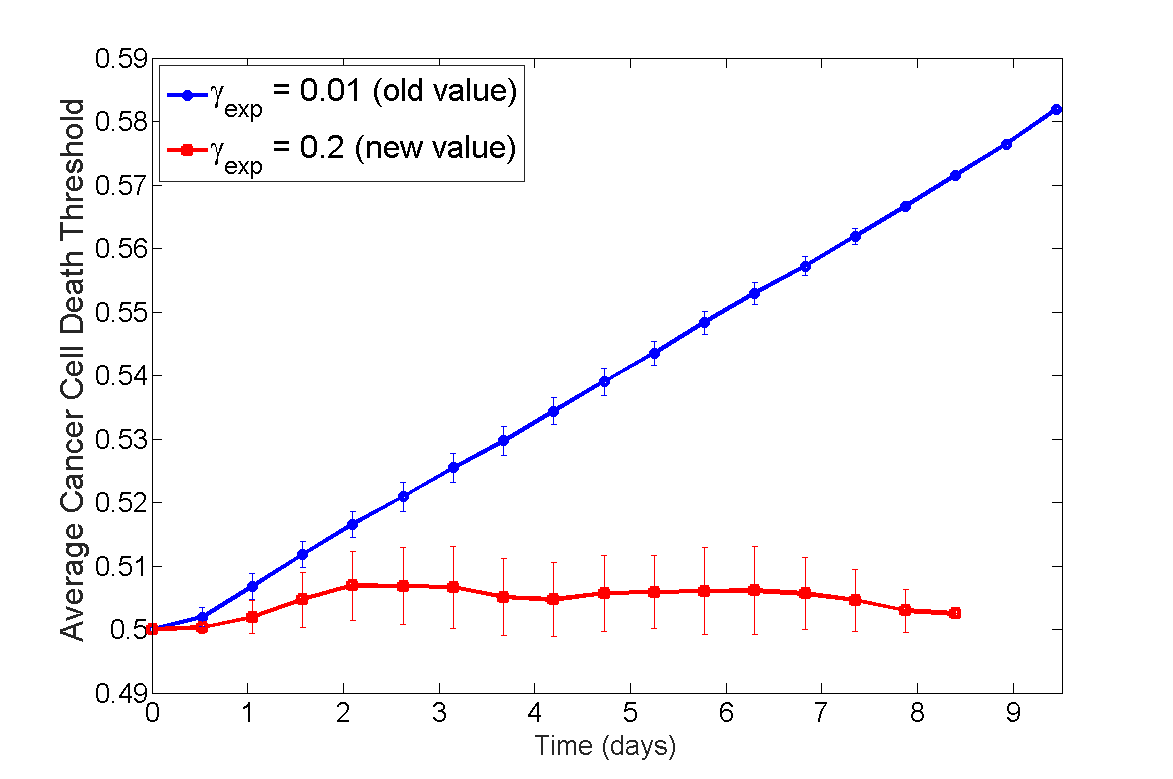}\\
  \caption{\footnotesize{Plot of average death threshold as a function of time using $\gamma_{exp} = 0.01$  as done in \cite{Gevertz15} (blue circles), and using $\gamma_{exp} = 0.2$ as done herein (red squares). Both simulations are metronomic therapy in the case of only acquired resistance with $\Delta_{death} = 2\times 10^{-5}$ and $p = 3\times 10^{-4}$. Note the linear behavior of the average death threshold with a very small standard deviation when $\gamma_{exp}=0.01$ is indicative that almost all cells are meeting the criteria of enough drug concentration for long enough time to acquire resistance. Using $\gamma_{exp} = 0.2$ adds variability in the average death threshold, meaning not all cells are meeting the acquired resistance criteria at all points in time.}}
  \label{figure:exposure_conc}
  \end{center}
\end{figure}

In \cite{Gevertz15}, $\gamma_{exp}=0.01$, which was such a low concentration of drug that for most of time, all the cells in space had drug concentrations greater than $\gamma_{exp}$. This created a situation where, on average, the death threshold of all cells increased linearly as a function of time with slope of $\gamma_{exp}$ (see Fig. \ref{figure:exposure_conc}). The lack of  spatial variability in terms of how cells acquire resistance to the drugs is further illustrated when observing the small standard deviation in the death threshold for this value of $\gamma_{exp}$. In order to allow for more variability in the acquired case, in this work we use a concentration threshold twenty times larger ($\gamma_{exp}=0.2$), so that cells with more access to the drug develop resistance quicker than cells with less access. This adds variability to the otherwise linear growth of the death threshold as a function of time (Fig. \ref{figure:exposure_conc}).

\section{Results and Discussion}

Our goal in this work is to explore the effects the dosing schedule of a DNA damaging drug has on the development of chemoresistance in simulated tumor micrometastases. Each simulation will start from a small group of cancer cells embedded in the tissue slice (as shown in top of the left column in Fig. \ref{figure:flowchart}). We will separately consider two kinds of anti-cancer resistance: acquired or pre-existing, and for each of them will examine two hypothetical cancer cell lines that differ in their pharmacological response to the drug under consideration. In order to evaluate the different drug administration schedules, we will compare average tumor response over ten replicates per treatment protocol. 

Our baseline treatment will be a metronomic, continuous infusion of drug (model parameter $S_{\gamma}= 1$ in each iteration, with one day corresponding to 960 iterations). Following the experimental data on ovarian cancer xenografts \cite{DeSouza11}, we will consider weekly treatment protocols in which on each day of the week, drug is either administered continuously or not given (fractionated-dose (FD) therapy). However, to constrain the overall drug toxicity over a period of a week, we will control the amount of drug administered during that time. That is, at the end of the week, no matter how many days the treatment was given, each mouse in \cite{DeSouza11} received the same amount of drug, and therefore by the end of each week in our simulations, the total amount of drug administered is fixed. Under this constraint, the supply rate of drug for a protocol is given by $\left(7/N\right)S_{\gamma}$, where $N$ is the number of days drug is given in a week. 

\begin{figure}[ht!]
 \begin{center}
  \includegraphics[width=0.95\textwidth]{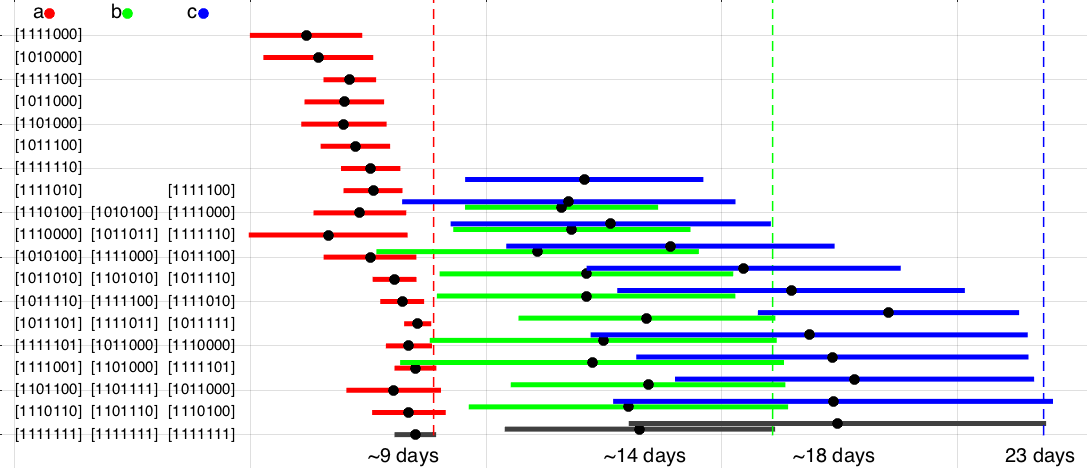}\\
  \caption{\footnotesize{Top treatment protocols for micrometastases with (a) weakly acquired, (b) weakly pre-existing and (c) strongly acquired resistant cells. The average time of eradication for each treatment is denoted by a circle; the standard deviation taken over ten runs of the same protocol is denoted as a horizontal line. Metronomic therapy in each case is shown on the bottom as a grey line and the corresponding average time to eradication is listed. For each treatment schedule: 1 indicates that drug is administered and 0 indicates no drug is administered on a given day of a week. Note, if a tumor was not eradicated, the ``time to eradication" for that realization of the protocol was set to 21 days, the length of the treatment window.}}
  \label{figure:rank_protocols}
  \end{center}
\end{figure}

As in \cite{DeSouza11}, the one-week treatment protocol will be administered three times, for a total of three weeks of treatment. Using our model, we will test all $2^7 - 1= 127$ possible treatment protocols of either giving (indicated with a 1) or not giving (indicated with a 0) drug on each of the seven days of the week, leaving out the no treatment case [0000000] when no drug is given every day of the week. Note that our computational framework allows us to test many more schedules than the three tested in the experimental work in \cite{DeSouza11}: continuous, three injections per week, one injection per week. This allows us to explore a much larger scheduling-space than can be feasibly done experimentally. The top-performing protocols identified are summarized in Fig. \ref{figure:rank_protocols}, with more details given in the subsections that follow.

\subsection{Acquired Resistance}

Here we use the previously-described model with the probability of DNA damage repair fixed at $p = 3\times 10^{-4}$. In \cite{Gevertz15}, we identified that a tumor treated with metronomic chemotherapy can fall in one of three parameter regimes: tumor eradication (weak resistance), initially responsive tumor that becomes drug resistant (strong resistance), complete treatment failure (no measurable response to metronomic protocol). For further consideration, here we will study two cells lines, dismissing those in the final parameter regime, as drugs which never cause a decrease in tumor size in the model likely have parameter values outside the scope of clinically-approved drugs. The first cell line we will study is a weakly resistant one that can be eradicated by metronomic therapy: $\Delta_{death}$=$7\times 10^{-5}$. The second is a strongly resistant cell line that cannot be eradicated by metronomic therapy: $\Delta_{death}$=$1.2\times 10^{-4}$.

\subsubsection{Optimization for Weakly Resistant Micrometastases}

We performed a detailed analysis of drug administration protocols on a cell line for which the rate that DNA damage increases due to prolonged drug exposure is given by $\Delta_{death} = 7\times 10^{-5}$. Our goal was to find treatment protocols that are at least as effective as metronomic chemotherapy, which we defined using the following two criteria:
\begin{itemize}
 \item Time of tumor eradication (average time + standard deviation taken over ten protocol replicates) does not exceeds  the time of tumor eradication for the metronomic schedule (again, average + standard deviation over ten schedule replicates) by more than 3\%,
 \item Protocol eradicates at least 30\% of micrometastases (30\% was determined based on the observation that the metronomic schedule for strongly acquired resistance has 30\% effectiveness).
\end{itemize}

\begin{figure}[ht!]
 \begin{center}
  \includegraphics[width=0.95\textwidth]{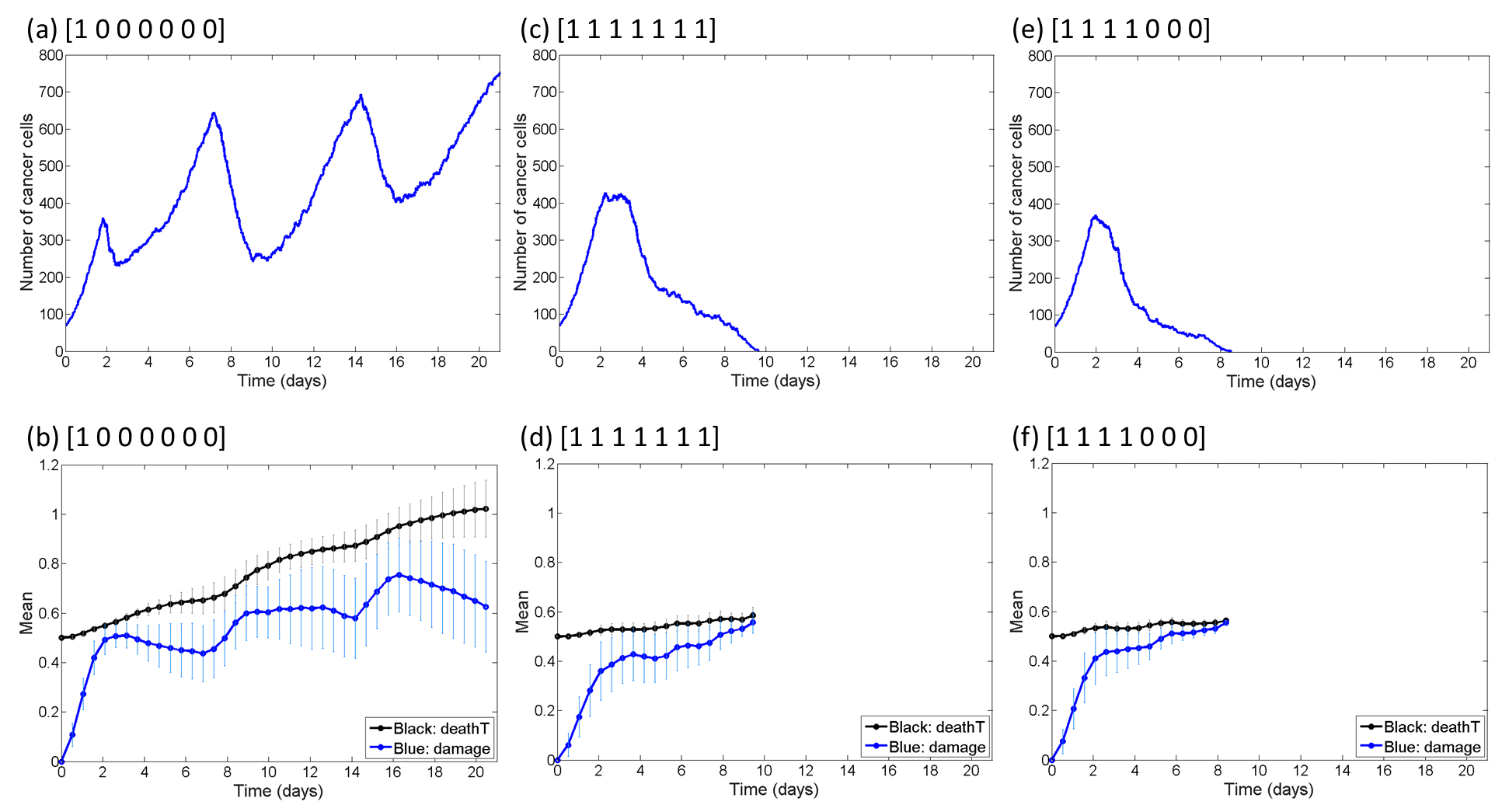}\\
  \caption{\footnotesize{Efficacy of various treatment protocols in the case of micrometastases with weak acquired resistance. Top row shows number of cancer cells over time, and bottom row shows the dynamics of the average death threshold of the cancer cells ($C_k^{death}$, called `deathT' in plots), as well as the average damage levels ($C_k^{dam}$, called `damage' in plots). (a)-(b) represents a MTD protocol that is one of the worst-case scenarios; (c)-(d) represents MC; (e)-(f) represents the optimal treatment protocol, a FD protocol that on average eradicates tumors the fastest.}}
  \label{figure:acquired_eradication3}
  \end{center}
\end{figure}

Fig. \ref{figure:rank_protocols}(a) shows schematically all protocols that either outperformed, or performed just as well as metronomic therapy for weakly resistant micrometastases. 18 of the 126 non-MC protocols tested met these criteria, and all 18 protocols were successful at eradicating 100\% of micrometastases (10 different seeds for the random number generator). According to Fig. \ref{figure:rank_protocols}(a), the protocols that on average outperform or perform on-par with MC all require that drug be administered on day one with 1-4 off-days during the week. These off-days are often found in the later part of the week: 72\% of off-days in the protocols that perform at least as well as MC fall within days 5-7 of the week. However, if the drug dose is too high at the beginning of the week and the off-time is too long (such as in protocol [1000000] shown in Fig. \ref{figure:acquired_eradication3}(a)-(b), an example of the MTD protocol), the treatment fails to eradicate a tumor in any of the ten trials conducted. In other words, FD protocols in which moderate doses of drug are given at the beginning of the week appear to be the optimal way to target a tumor with only acquired resistance, at least in the weak resistance regime.

The result of treating a tumor with any one of these top protocols follows the trend shown for the optimal protocol [1111000] (optimal in that it gives the quickest average time to eradication and has the smallest value of average time to eradication plus standard deviation) in Fig. \ref{figure:acquired_eradication3}(e)-(f). We find the dynamics in this case are very similar to the case of metronomic therapy (Fig. \ref{figure:acquired_eradication3}(c)-(d)): the average damage level rapidly catches up to the average death threshold, resulting in tumor eradication. The optimal strategy is superior to metronomic therapy in that the the maximum size the tumor can achieve is smaller when giving the FD protocol, presumably because of the higher drug levels cells are exposed to in the beginning days of treatment.

Simulations also revealed that 81 of 127 treatment protocols were able to eradicate the weak acquired micrometastases 100\% of the time, meaning there are many different treatment protocols that would lead to tumor eradication in this parameter regime. Therefore, it is also interesting to explore the strategies that cannot eradicate the tumor, and try to understand what goes wrong in those cases. These strategies have at least two of the following common features: 1) Relatively high doses of the drug administered each day it is given; 2) Limited amount of drug given on the first 4 days of treatment (0-1 doses), allowing the tumor ample time to grow before significant amounts of drug are given; 3) Large spacing between drug doses. The MTD schedule of [1000000] discussed above is explored further in Fig. \ref{figure:acquired_eradication3}(a)-(b). In this case, the tumor oscillates wildly in size due to the high drug dose given and the long rest period. Since the average death threshold of the cancer cells increases, while the average damage level is actually decreasing, continued administration of this weekly dosage schedule is not expected to ever result in tumor eradication.


\subsubsection{Optimization for Strongly Resistant Micrometastases}

In this parameter re\-gime, we consider a cell line for which the rate of DNA damage increase due to prolonged drug exposure is given by $\Delta_{death} = 1.2\times10^{-4}$. Since the MC protocol fails to eradicate the majority of tumors (fails 70\% of the time) for this parameter value, here we will search for more effective protocols. 

\begin{figure}[h!]
 \begin{center}
  \includegraphics[width=0.95\textwidth]{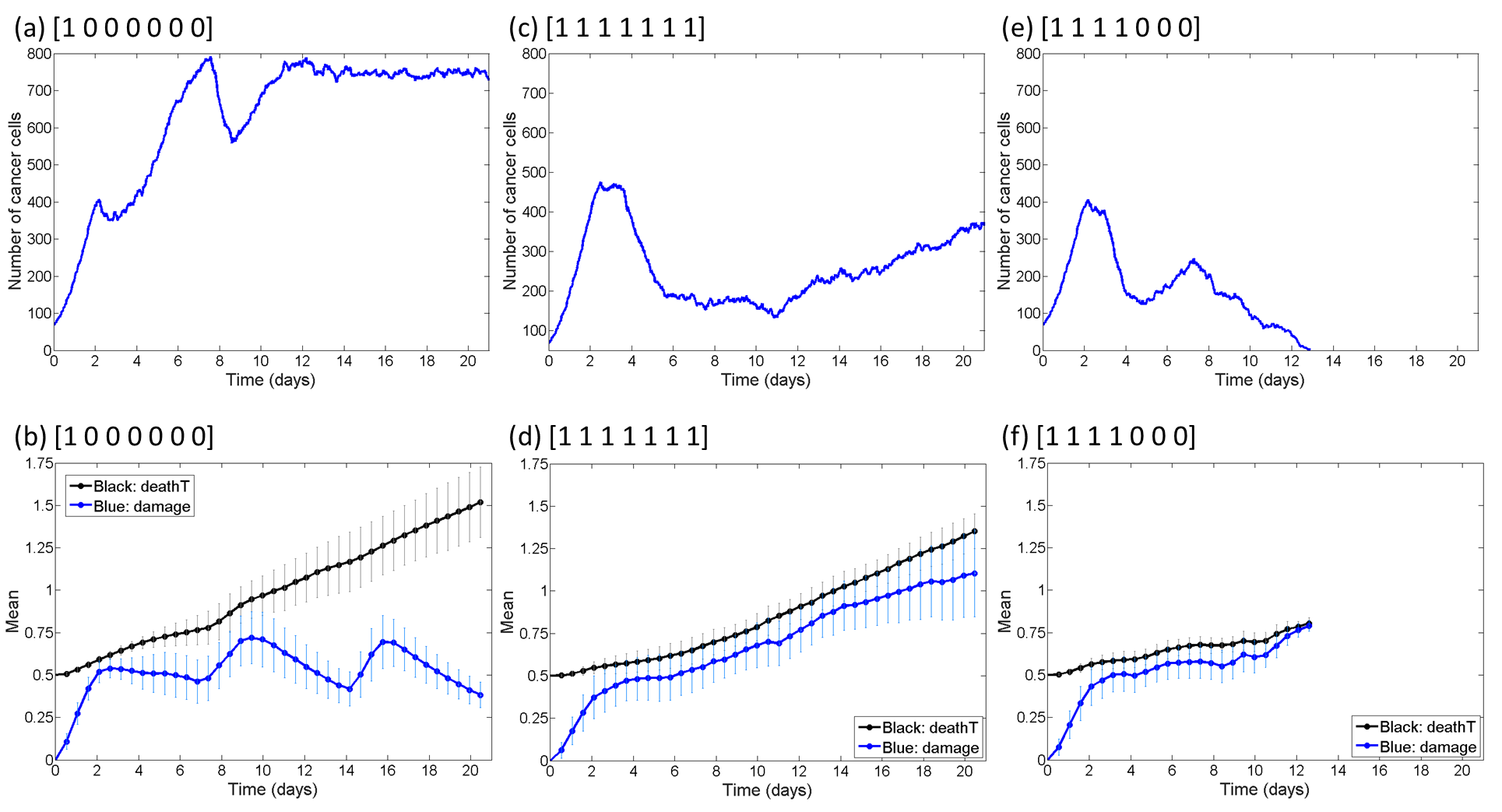}\\
  \caption{\footnotesize{Efficacy of various treatment protocols in the case of micrometastases with strong acquired resistance. Top row shows number of cancer cells over time, and bottom row shows the dynamics of the average death threshold of the cancer cells ($C_k^{death}$, called `deathT' in plots), as well as the average damage levels ($C_k^{dam}$, called `damage' in plots). (a)-(b) represents an MTD protocol that fails 100\% of the time; (c)-(d) represents an instance in which MC fails to eradicate the tumor, which happens approximately 70\% of the time; (e)-(f) represents an optimal treatment protocol, a FD protocol that, on average, results in tumor eradication in the quickest time.}}
  \label{figure:acquired_drug_resist3}
  \end{center}
\end{figure}

We observe that it is much harder to find treatment protocols to eradicate micrometastases with strong resistance. Only 16 of the 127 treatment protocols tested were successful at tumor eradication in this parameter regime for at least 1 of the 10 protocol replicates.  11 protocols were found to meet the standard for performing at least as effectively as MC, with only six of these resulting in tumor eradication in more than the majority of replicates. 10 of the 11 protocols that are at least as effective as MC in the case of micrometastases with strong acquired resistance are also at least as effective as MC in the case of weak acquired resistance (see Fig. \ref{figure:rank_protocols}(a) versus (c)). This strongly suggests that the top strategies for acquired resistance do not greatly depend on whether resistance is weak or strong. 

Another similarity between the weak and strong acquired resistance cases is that any protocol that fails on micrometastases with weakly resistant cells also fails on those with strongly resistant cells. For instance, returning to the MTD protocol [1000000], simulations reveal that high drug levels at the beginning of treatment cause a strong acquired micrometastases not to be eradicated. The very high dose drives resistance too quickly (Fig. \ref{figure:acquired_drug_resist3}(b)), and a tumor cannot recover from this during the break (Fig. \ref{figure:acquired_drug_resist3}(a)).

To further compare the response of weak and strong acquired micrometastases, we observe that the top protocol in the case of weak acquired resistance ([1111000]) is also the top protocol in the case of strong acquired resistance in terms of average time to eradication (it is the second best in terms of average time plus standard deviation). In spite of these similarities with the weak acquired case, there is a stark difference in response to MC when comparing the weak (Fig. \ref{figure:acquired_eradication3}(b)) and strong (Fig. \ref{figure:acquired_drug_resist3}(b)) acquired cases. In particular, MC has only a 30\% chance of tumor eradication in the strongly acquired parameter regime, whereas the optimal protocol has a 90\% chance. Compare this to the weakly acquired case in which both protocols result in 100\% eradication.


\subsection{Pre-Existing Resistance}

Using the previously-described model and fixing the multiplier that determines how much more damage the resistant cells can tolerate compared to sensitive cells ($Thr_{mult} = 3.25$), we allowed the probability of DNA damage repair $p$ to vary. Just as in the case of acquired resistance, pre-existing resistant micrometastases treated with metronomic chemotherapy can fall in one of three parameter regimes, depending on the value of $p$: weak resistance, strong resistance, complete treatment failure \cite{Gevertz15}. For further consideration, we will study micrometastases compromised of approximately 3\% pre-existing resistant cells from one of the following two parameter regimes: weakly resistant  ($p = 6\times 10^{-5}$), and strongly resistant ($p = 3\times 10^{-4}$).


\subsubsection{Optimization for Weakly Resistant Micrometastases}

This parameter regime ($Thr_{mult} = 3.25$, $p = 6\times 10^{-5}$) was classified as one in which the MC protocol can eradicate the tumor in spite of the presence of pre-existing resistant cells. In this section, we sought to find alternative treatment protocols that could eradicate micrometastases at least as effectively as MC. We found that 10 of the treatment protocols tested out-performed or performed on par with MC (Fig. \ref{figure:rank_protocols}(b)). Each of those protocols  were successful at eradicating the tumor for each of the 10 replicates. Similar to what was observed for successful protocols in the case of weak acquired resistance, each of the protocols that perform at least as well as MC initiate treatment on day one. Further, they have 1-4 off-days scattered throughout the week, with over 69\% of the off-days being in the latter half of the week (days 5-7). Half of the top protocols for micrometastases with weak pre-existing resistance were also top protocols for micrometastases with weak acquired resistance (compare Fig. \ref{figure:rank_protocols}(a) to (b)).

\begin{figure}[ht!]
 \begin{center}
  \includegraphics[width=0.95\textwidth]{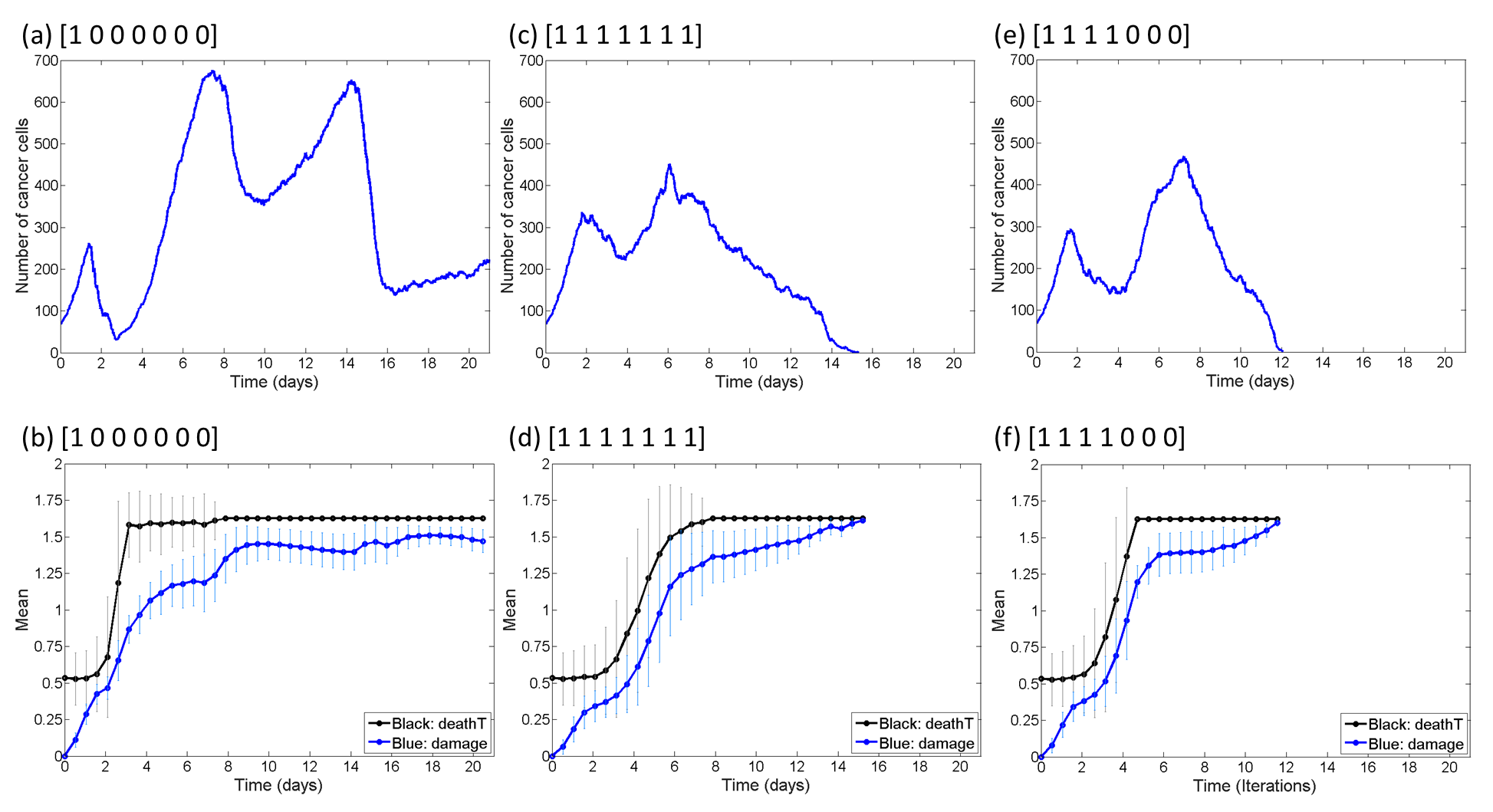}\\
  \caption{\footnotesize{Efficacy of various treatment protocols in the case of micrometastases with weak pre-existing resistance. Top row shows number of cancer cells over time, and bottom row shows the dynamics of the average death threshold of the cancer cells ($C_k^{death}$, called `deathT' in plots), as well as the average damage levels ($C_k^{dam}$, called `damage' in plots).  (a)-(b) represents a MTD protocol that fails 100\% of the time; (c)-(d) represents MC; (e)-(f) represents the optimal FD protocol, in terms of average time to eradication (third best in terms of average time plus standard deviation).}}
  \label{figure:pre_drug_erad3}
  \end{center}
\end{figure}

The result of treating a tumor with any one of these top protocols follows the trend shown in Figure \ref{figure:pre_drug_erad3}(e)-(f). The resistant cell population is always selected for, as evidenced by the average value of the death threshold plateauing at 0.5$\times$3.25 = 1.625 (the value of the death threshold in the resistant cell population), with the standard deviation decreasing to zero as the plot plateaus (meaning all cells have this higher death threshold and the sensitive cells have been eliminated from the population). In spite of the fact that the resistant cells are selected for, the damage induced by the drug is not repaired quickly enough, and eventually the resistant cells are killed by the drug.

The reason this FD protocol is superior to MC is that the sensitive population of cells is eradicated quicker. In the protocol realization shown in Fig. \ref{figure:pre_drug_erad3}(f), it takes under 5 days for optimal FD protocol to eradicate the sensitive tumor cells, while it takes nearly 8 days for MC to eradicate these cells (Fig. \ref{figure:pre_drug_erad3}(d)). The quicker removal of the sensitive cells in the case of the optimal FD protocol allows more drug access to resistant cells, which results in a quicker build-up of damage, and hence a quicker time to elimination of the resistant population (compare Fig. \ref{figure:pre_drug_erad3}(e) to (c)).

It is of note that all but 8 of the 127 treatment protocols tested can eradicate micrometastases with weak pre-existing resistance for at least 1 of the 10 replicates, and 28 of 127 eradicated these micrometastases in each of the 10 trials. In other words, there are many treatment protocols that can lead to tumor eradication. Since so few treatment protocols fail, it is interesting to explore what make a schedule unsuccessful. Figure \ref{figure:pre_drug_erad3}(a)-(b) looks at one such case: the MTD protocol [1000000]. Surprisingly, we see that the sensitive cells have been eradicated by this MTD protocol in just under 8 days, quite comparable to what is observed for MC. Yet, the removal of the sensitive cells is not a sufficient condition for a protocol to be successful, as the damage level has not surpassed the death threshold in well over 200 cells (over 20\% of the domain size) by the end of the treatment window. The large number of off-days in this protocol, coupled with the very high dose of drug given on the on-day, leaves a number of cells accumulating damage too slowly to be eradicated by this protocol.


\subsubsection{Optimization for Strongly Resistant Micrometastases}

In the case of mi\-cro\-me\-ta\-sta\-ses composed of 3\% strongly pre-existing resistant cells, none of the 127 treatment protocols tested are capable of causing tumor eradication in any of the replicates. The resistant subpopulation is simply too tolerant of DNA damage for the damage level of these cells to surpass the death threshold at the constrained weakly dose of drug. The fact that (in certain parameter regimes), the presence of a small number of pre-existing resistant cells can prevent tumor eradication (with a high probability) has also been concluded in other modeling studies; see for instance \cite{Komarova07,Foo09,Menchon15}.


\subsection{Effective and Ineffective Treatment Protocols}

We previously discussed treatment schedules that performed at least as well as metronomic therapy for three types of micrometastases: weak acquired resistance, strong acquired resistance, weak pre-existing resistance (Fig. \ref{figure:rank_protocols}). However, in the case of weak acquired and weak pre-exisitng resistance, we also identified several additional schedules that have high probability of tumor eradication, but require more time than the metronomic protocol to eliminate the tumor. Such schedules are effective, but not as efficient as metronomic therapy.

\begin{figure}[h!]
\begin{center}
  \includegraphics[width=0.9\textwidth]{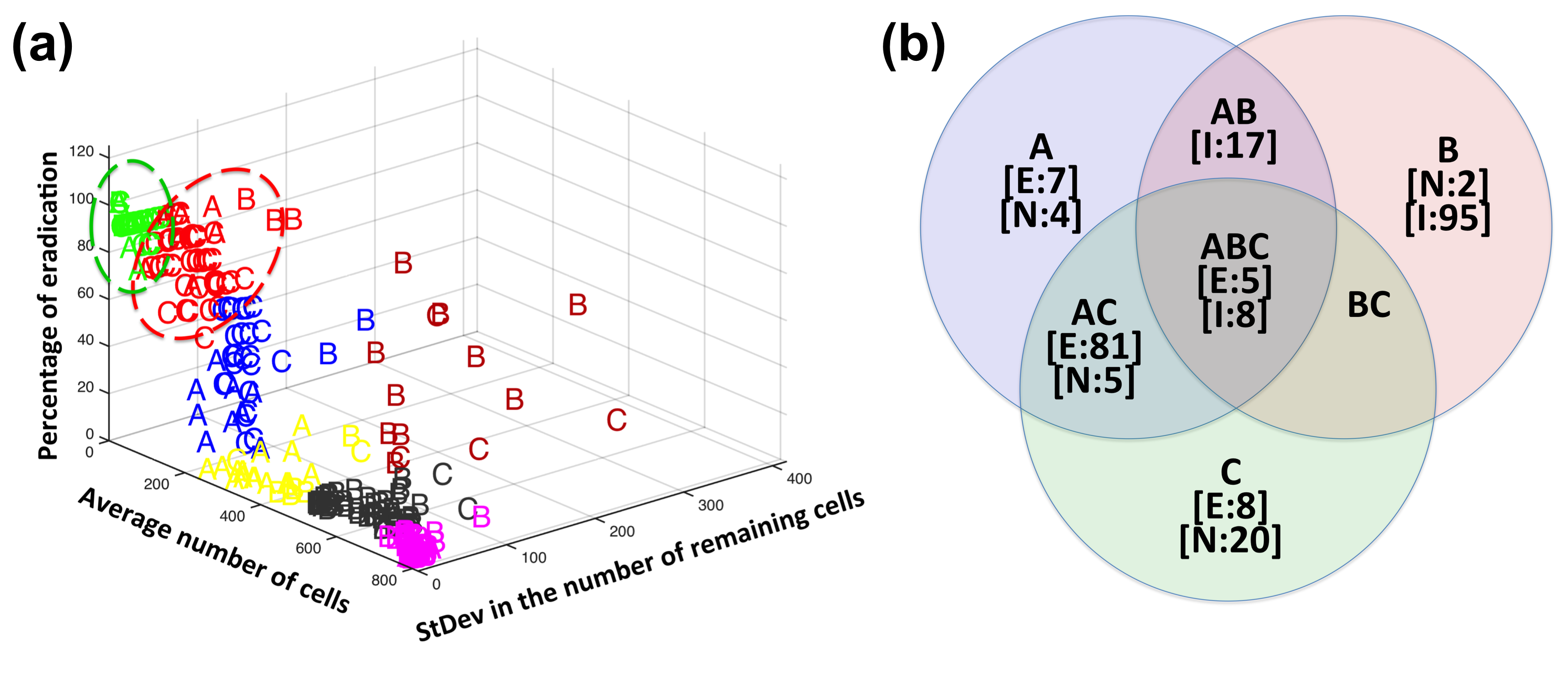}
  \caption{\footnotesize{ (a) Partition of all simulated protocols for three cell lines (A-C) into clusters based on three metrics: the average number of remaining tumor cells, their standard deviations and the percent of eradicated tumors. Two treatments clusters (circled) have been classified as effective, one cluster (blue) as neutral, and the remaining four clusters as ineffective. The effective green cluster contains 139 elements centered around the following treatment outcomes: no surviving cancer cells, and 100\% eradication potential. The effective red cluster contains 53 elements centered around the following treatment outcomes: tumor size of 33.4 cells, a standard deviation of 66.9 cells, and an eradiation potential of 70\%. (b) A three-set Venn diagram showing numbers of effective [E:], [N:] neutral, and ineffective [I:] protocols for all combinations of three considered cell lines (zero-counts are omitted). The three tumor cell lines considered are: A--weak acquired resistance; B--strong acquired resistance; C--weak pre-existing resistance. }}
  \label{figure:VennDiagram}
  \end{center}
\end{figure}

Here, we used data clustering techniques to partition all protocols into separate groups (clusters) and to compare protocol effectiveness between different cell lines. Using a $k$-medians clustering algorithm, we divided all data into groups presented in Fig. \ref{figure:VennDiagram}(a) taking into account the following three measurements: (i) the average (over ten treatment replicates) number of cells that have not been eradicated by the three-week protocol, (ii) the standard deviation in the remaining cell number, and (iii) the percentage of treatments that led to tumor eradication (eradication potential). Each of the seven generated clusters are concentrated near the cluster median (a cluster center) that serves as a prototype of that cluster.

The first cluster to be classified as effective (green in Fig. \ref{figure:VennDiagram}(a)) has a center representing the ideal treatment protocol: no remaining cells, zero standard deviation and 100\% eradication. The only other cluster we classified as effective (red in Fig. \ref{figure:VennDiagram}(a)) is characterzied by the following center: an average of 33.4 surviving cancer cells, a standard deviation of 66.9 cells, and an eradication potential of 70\%. The remaining clusters, with the exception of the blue cluster, are considered ineffective as they either result in large tumors at the end of the treatment window, or have very low eradication potential (less than 30\%). The blue cluster appears to be an intermediate (neutral) case between the effective and ineffective protocols. The blue cluster center represents tumors with an average size after treatment of 105.8 $\pm$ 87.3 cells (accounts only for about 10\% of the domain space), and an eradication potential of 30\% (same potential we observed for MC in the case of strong acquired resistance). The ineffective clusters have cluster centers representing significantly larger tumors, and less eradication potential.

The numbers of effective and ineffective protocols for each of the three cell lines are shown in the Venn diagram in Fig. \ref{figure:VennDiagram}(b), where the non-zero counts of effective [E:], neutral [N:] or ineffective [I:] protocols are shown for each of the possible cell line combinations. All 127 protocols, with ten replicates per protocol, are represented in the Venn diagram for each of the three cell lines. We found substantial overlap between effective protocols for the weaky resistant cells: 81 of the 127 protocols were effective at eradicating micrometastases with weakly acquired and weakly pre-existing resistant cells (see case AC in Fig. \ref{figure:VennDiagram}(b)). Surprisingly, there were no protocols effective in both the weakly and strongly resistant cases that were also not effective for weakly pre-existing resistant cells (see AB in Fig. \ref{figure:VennDiagram}(b)).

The Venn diagram classification of treatment protocols also allows us to identify a set of schedules of high effectiveness that one can choose from in designing a personal treatment plan, independent of the type of resistance a patient's micrometastases may harbor. We identified five protocols ([1111000], [1111100], [1111110], [1011100], [1011110]) that are classified as ``effective" at eliminating all three types of tumors with very high probability (see the case ABC of Fig. \ref{figure:VennDiagram}(b)). This subset of protocols represent potentially powerful dosing schedules for targeting micrometastases composed of cells with differing resistance types and capacities. Among the subset of five treatments, two fractionated-dose protocols ([1111000] and [1111100]) outperform metronomic therapy in all three cell lines considered (see Fig. \ref{figure:rank_protocols}). The robust performance of these protocols over a range of resistance types and treatment parameters suggests they may be the dosing schedules with the highest likelihood of eradicating micrometastases from cells with differing (and generally, unknown) resistance types and capacities.


\subsection{The Role of Tumor Microenvironment}

\begin{figure}[ht!]
\begin{center}
  \includegraphics[width=0.8\textwidth]{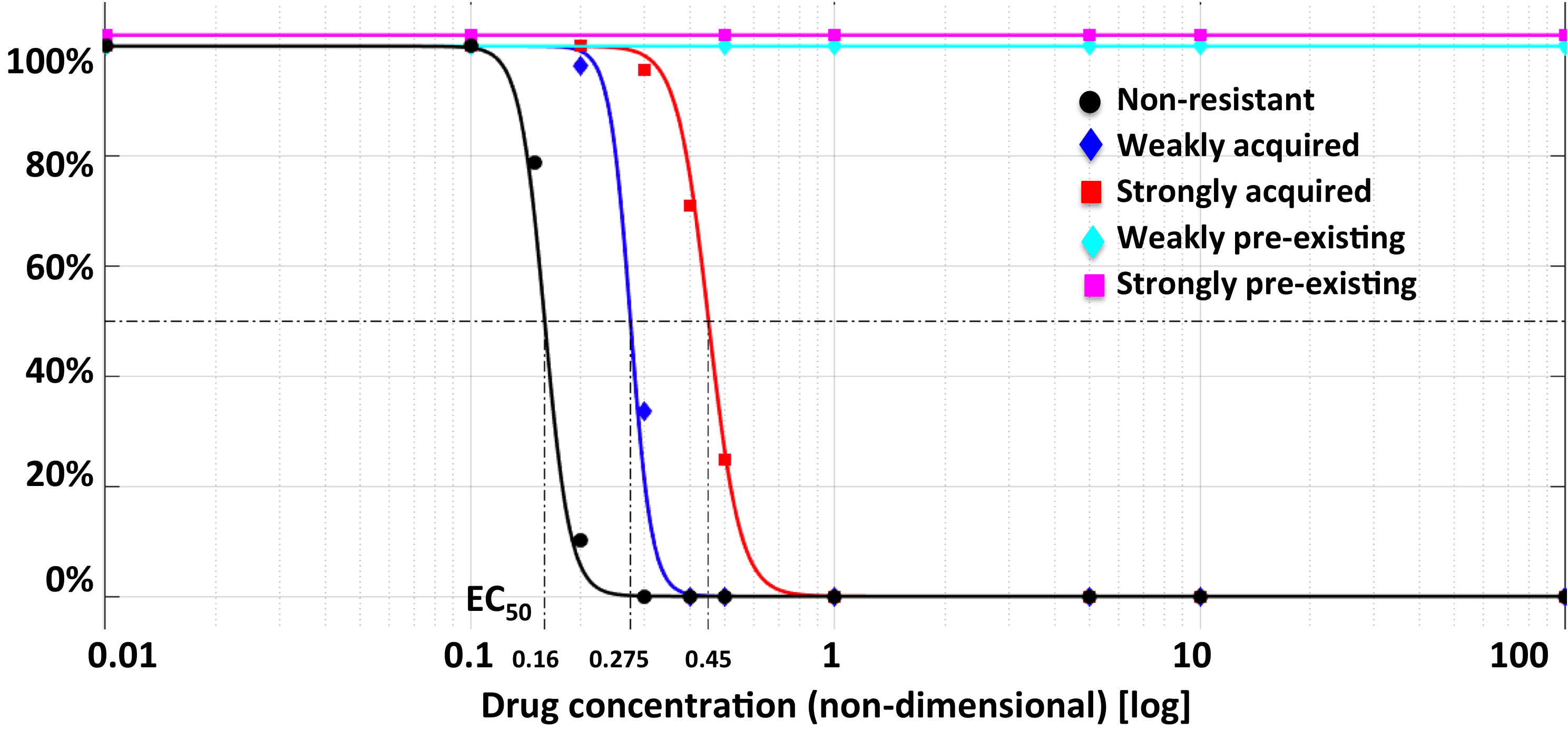}
 \caption{\footnotesize{$EC_{50}$ curves of cancer cell response to the simulated DNA damaging drug. Cell lines shown are: a non-resistant (black) with an $EC_{50}$ value of 0.16, weak acquired resistance with $\Delta_{death} = 7\times 10^{-5}$ (blue), strong acquired resistance with $\Delta_{death} = 1.2\times 10^{-4}$ (red), weak pre-existing resistance with $p = 6\times 10^{-5}$ (cyan), and strong pre-existing resistance with $p = 3\times 10^{-4}$ (magenta).} The $EC_{50}$ curves were generated using the 2D cell colony simulations in which a one-time injection of drug was given, and cellular response was recorded for 72 hours. The fitting curves satisfy equation: $y(x)=min + (max - min)/(1+ (\frac{x}{EC_{50}})^{-\beta})$, where $min$ and $max$ are the lowest and the highest observed value. $\beta$, which characterizes the steepness of the curve at its midpoint, is the so called Hill slope of the four parameter logistic (4PL) nonlinear regression model.}
\label{figure:EC50}
\end{center}
\end{figure}

The importance of he\-te\-ro\-ge\-neities in the tumor microenvironment, such as an irregular tumor vasculature which results in the formation of drug and metabolite gradients, can be illustrated by comparing how the four tumor cell lines respond to homogeneous vs. heterogeneous drug concentrations. Typically, the drug response curves ($IC_{50}/EC_{50}$) are generated by growing the tumor cells in monolayers (in Petri dishes) for 72 hours. While a uniform drug concentration is applied to each dish, it progressively increases between the dishes \cite{TurnerCharlton:2005}. The $EC_{50}$ value is defined as the drug concentration that gives half-maximal response (growth inhibition for the $IC_{50}$) after a specified exposure duration. In our simulations we record the number of cells that have survived a one time dose into a simulated 2D cell culture at various drug concentrations. The $EC_{50}$ curves for all four cell lines considered in our study, as well as for the chemosensitive cell line, are presented in Fig. \ref{figure:EC50}.

In the case of pre-existing resistance, both the weak and strong resistance cell lines have no defined $EC_{50}$ value, since there is no decrease in their population cell counts, even for extremely high drug concentrations. Yet, MC was able to eradicate a micrometastasis containing approximately 3\% of the weak pre-existing resistant cells, contrary to what was shown in the $IC_{50}$ experiments. This failure to differentiate between these cell lines can be attributed to the short time of the Petri dish experiments compared to the much longer time used in the tissue-based simulations. 

In the case of acquired resistance, the weakly and strongly resistant cell lines have $EC_{50}$ values that are on the same order of magnitude: the $EC_{50}$ value for the weakly resistant cell line is 0.275, while the value for the strongly resistant cell line is 0.45. Despite having $EC_{50}$ values on the same order or magnitude, these cell lines respond very differently in our tissue-based simulations. In the cell-culture simulations, all cells (independent of their position in space) have equal exposure to the drug. On the other hand, in the tissue-based simulations, cells are exposed to a variable drug gradient generated by irregularly spaced vessels. The impact of the variable drug gradient is more pronounced for the cell line with strongly acquired resistant cells. As we showed in \cite{Gevertz15}, the heterogeneous vascular architecture we are considering produces microenvironmental niches that promote the formation of drug resistance, particularly when that resistance is sufficiently strong as it is in the strong acquired resistance parameter regime. Since no such niches exist in cell culture, the cell culture experiments cannot distinguish between which cell lines will respond to MC in the tissue simulations, and which will not.


\section{Conclusions and Future Directions}

In this paper we examined the impact that the dosing protocol for systemic chemotherapy has on the development of drug resistance at metastatic sites. This study utilized a spatial, agent-based model of micrometastatic growth in heterogeneous microenvironments. We showed previously that two specific microenvironmental niches -- the niche with low oxygen content, and the niche with low drug levels but sufficient oxygen concentration -- form sanctuaries in which tumor cells having a chemoresistant potential can give rise to a resistant tumor \cite{Gevertz15}. Here, four distinct tumor cell populations, each categorized by their response to a DNA damaging drug, were considered: 1) a population that can weakly acquire resistance; 2) a population that can strongly acquire resistance; 3) a population composed of 3\% weakly pre-existing resistant cells; 4) a population composed of 3\% strongly pre-existing resistant cells. The weakly resistant cell lines were defined by those that could be eradicated by a metronomic (daily, low-dose) treatment schedule.

Focusing on the two weakly resistant micrometastases, 50\% of the protocols that are at least as effective as metronomic therapy in the case of pre-existing resistance are also at least as effective in the case of acquired resistance (Fig \ref{figure:rank_protocols}). If we then classify the treatment protocols as effective or ineffective (without concern for if they outperform MC), we further observe significant overlap between the effective protocols in these instances (Fig. \ref{figure:VennDiagram}). Expanding our analysis to include micrometastases with strongly acquired resistant cells, we found that approximately 90\% of the protocols that are at least as effective as metronomic therapy in the case of strong resistance are also at least as effective in the case of weak resistance. This suggests that the top protocols for targeting micrometastases that can acquire resistance to a DNA damaging drug are not strongly dependent on the rate at which these cell acquire resistance.

Further, any protocol that is effective for both the weakly and strongly acquired cases is also effective for micrometastases with weak pre-existing resistance (in Fig. \ref{figure:VennDiagram}(b), AB has no effective treatments, but ABC has 5). This result was quite surprising, as in previous work we showed that the dynamics of pre-existing and acquired resistant cell lines vary in significant ways, with acquired resistance dynamics being highly dependent on the heterogeneous microenvironment, and pre-existing resistance dynamics being mainly driven by the inherent genetic/epigenetic advantage of the resistant cells \cite{Gevertz15}.

It is of interest to compare model findings to the experimental work of De Souza and colleagues in which several chemotherapy administration schedules were examined using murine xenografts of a drug resistant ovarian cancer cell line \cite{DeSouza11}. In that work, they found that continuous treatment with docetaxel for three weeks resulted in significantly decreased tumor burden compared to mice treated intermittently (with a protocol that either gives drug once or three times per week) \cite{DeSouza11}. From one perspective, this stands in contrast to our findings that several fractionated (intermittent) protocols could perform at least as well as, if not better, than MC in a number of simulated cell lines. However, it is worth noting that our computational study examined 127 protocols, meaning there is much more opportunity to find optimal protocols when compared to an experimental study only examining three schedules. That said, similar to the work in \cite{DeSouza11}, we did find significant anti-tumor benefits of continuous therapy over any protocol that administers drug once per week. And, although in \cite{DeSouza11} no benefit was found in giving drug three times per week, our computational analysis revealed that the anti-tumor activity of the majority of protocols that give drug three times per week (31 of 35 protocols) is not comparable or superior to MC for any simulated cell line tested in this study. In order to directly compare our results to those in \cite{DeSouza11}, we would need to know precisely which days of the week the three drug doses were given. That said, there could be other reasons for the (potential) discrepancy between the model and xenograft predictions using three days of treatment, including: 1) the microenvironment differences between a xenograft and a de novo tumor (which we are simulating), 2) the possibility that MC has different drug targets than MTD or other fractionated protocols, and we do not consider these different targets in our model.

Note that the conclusions drawn in this work are dependent on two key assumptions. First is that resistance is modeled as a \textit{neutral} mutation in the absence of drug (similar to the assumption in \cite{Komarova05}), meaning that sensitive and resistant cells behave the same way before treatment initiation. Second is that in the presence of drug, resistant cells are not assumed to have any fitness disadvantage; i.e., they do not proliferate slower than sensitive cells. This assumption is based on the fact that we are modeling a DNA damaging drug, and that cells with loosened DNA damage sensing mechanisms (say through the loss of p53) can bypass cell cycle checkpoints. Since the checkpoint phase is where division would likely be slowed down, and since we are assuming our resistant cells bypass these checkpoints, we do not assume that resistant cells have a fitness disadvantage in the presence of a DNA damaging drug.

In the future, we plan to extend our analyses to consider micrometastases that can harbor pre-existing resistance to a DNA damaging drug while simultaneously being able to acquire resistance in response to the drug. This will require developing a thorough understanding of tumor behavior in this two-dimensional parameter space (extent of pre-existing resistance versus acquired resistance), and searching for optimal protocols over different regions of this parameter space. We anticipate a more restricted set of treatments can lead to tumor eradication in this combined resistance case, and therefore we will likely have to extend the set of treatments tested to include protocols where drug dose is not equally distributed over the number of drug-administration days per week. While this would require us to move away from performing an exhaustive search of treatment space, it also expands the class of treatments we can consider to include chemo-switch protocols in which periods of high drug doses are followed by periods of low doses \cite{Pietras05}, or adaptive therapies in which the specific treatment timing and dosage are determined by tumor response to the previous drug dose \cite{Gatenby09}. 

The intention of the current work, coupled with the proposed extension, is to better understand treatment response of micrometastases growing in heterogeneous environments. Our results may have implications for how chemotherapy appointments are scheduled in the clinic, particularly when accounting for the various constraints in designing an optimal treatment schedule for a given patient. As just one example, often chemotherapy is not given during weekends to give the patients quality time with his/her family. Given such constraints in the number of days the treatment can be administered to an individual patient, our clustering scheme can help identify patient-specific protocols that are predicted to be ``effective", and can suggest protocols to avoid that are predicted to be ``ineffective". Focusing on ineffective treatments, we found that the maximum tolerated dose approach in which the weekly drug dose is given in one day is predicted to fail independent of type of resistance in the micrometatasis. The failure of MTD also held independent of which day of the week the drug was administered. On the other hand, five fractionated-dose protocols were classified as effective independent of the type and strength of resistance, and two of these ([1111000] and [1111100]) even out-performed metronomic therapy in all cases. This strongly suggests these protocols may be ideal ones to test in pre-clinical experimental studies with the eventual goal of implementing these in the clinic.



\section*{Acknowledgments} This work was supported in part by the NIH U01-CA202229-01 grant to KAR.



\bibliographystyle{plain}
\bibliography{references_resubmit}

\medskip
Received xxxx 20xx; revised xxxx 20xx.
\medskip

\end{document}